\documentclass[a4paper,10pt]{article}

\usepackage{amsfonts}
\usepackage{amssymb}

\begin{document}
\title{Metric fluctuations and the Weak Equivalence Principle
}
\author{Ertan G\"okl\"u and Claus L\"ammerzahl\\
ZARM Universit\"at Bremen, Am Fallturm, 28359 Bremen, Germany}
\maketitle

\begin{abstract}
We describe space--time fluctuations by means of small fluctuations of the metric on a given background metric. From a minimally coupled Klein--Gordon equation we obtain within a weak-field approximation up to second order and an averaging procedure over a finite space--time scale given by the quantum particle in the non--relativistic limit a modified Schr\"odinger equation. The dominant modification consists in an anomalous inertial mass tensor which depends on the type of particle and on the fluctuation scenario. The scenario considered in this paper is a most simple picture of spacetime fluctuations and gives an existence proof for an apparent violation of the weak equivalence principle and, in general, for a violation of Lorentz invariance.
\end{abstract}

\section{Introduction}
The unification of quantum mechanics and gravitation is one of the outstanding problems of contemporary physics. Though yet there is no theory of quantum gravity. However,  approaches like string theory or loop quantum gravity make some general prediction which may serve as guidance for experimental search. The difficulty is that in many cases the precise strength of the various expected effects are not known. Quantum gravity phenomenology tries to overcome this problem by parameterizing possible effects and to work out experimental consequences for classes of phenomena. Different quantum gravity scenarios are characterized by different parameters. One class of expected effects of a quantized theory of gravity is related to a nontrivial quantum-gravity vacuum ("spacetime foam'') which can be regarded as a fluctuating spacetime. 

Space--time fluctuations could lead to a minimal observable distance setting an absolute bound on the measurability of distances and defining a fundamental length scale \cite{{JaekRey1994PysLettA}, {KlinhamArxiv}}. For instance the search for additional noise sources in gravity-wave interferometers was considered \cite{GAC2000PRD}, which has also been analyzed in the context of an experiment with optical cavities \cite{SchillPRD2004}. A space--time foam may also violate the principle of equivalence as was suggested by Ellis and coworkers \cite{EllMavNanSak2004}.
Another prediction of quantum gravity models are so called deformed dispersion relations \cite{DeformedDispers0}-\cite{DeformedDispers3}.
In the work of Hu and Verdauger \cite{HuVerd2004LivRev} classical stochastic fluctuations of space--time geometry were analyzed stemming from quantum fluctuations of matter fields in the context of a semiclassical theory of gravity. This leads to a stochastic behaviour of the metric tensor.
Furthermore the effects of fluctuations of space--time geometry leading to, e.g.,  lightcone fluctuations, redshift and angular blurring were discussed in \cite{FordIntJTP2005}.
Space--time fluctuations can also lead to decoherence of matter waves which was discussed in \cite{PowerPercival1999} and \cite{Wang2006CQG}, where quantum gravitational decoherence is modelled by conformal space--time fluctuations.
The analysis in \cite{Camacho2003GRG} takes into account nonconformal fluctuations of the metric and yields a modified inertial mass which is subject to the stochastic properties.

In the present work we will also assume that such fluctuations shall manifest themselves as stochastic fluctuations of the metric. Generalizing the work \cite{Camacho2003GRG}, we will permit off-diagonal fluctuation terms in the metric.

We will show that our model of space--time fluctuations leads to a modified inertial mass which takes over the stochastic properties of the underlying fluctuation process and is dependent of the type of particle. It follows that necessarily an apparent violation of the weak equivalence principle occurs which is dependent of the noise model.

\section{Metrical fluctuations and noise properties}

We regard space--time as a classical background over which quantum Planck scale fluctuations are imposed, which appear as classical fluctuations of space--time. This assumption allows us to write a perturbed metric in the usual form 
\begin{equation}
g_{\mu\nu}(\mathbf{x},t) = \eta_{\mu\nu}+h_{\mu\nu}(\mathbf{x},t), \label{modmetric1}
\end{equation}
where $|h_{\mu\nu}| \ll 1$ and greek indices run from 0 to 3. The inverse metric is calculated to second order in the perturbations
\begin{equation}
g^{\mu\nu}(\mathbf{x},t) = \eta^{\mu\nu} - h^{\mu\nu}(\mathbf{x},t) + \tilde{h}^{\mu\nu}(\mathbf{x},t), \label{modmetric1a}
\end{equation}
with $\tilde{h}^{\mu\nu}=\eta_{\kappa\lambda}h^{\mu\kappa}h^{\lambda\nu}$. 
Indices are raised and lowered by means of the Minkowski metric $\eta_{\mu\nu}={\rm diag }(-1,1,1,1)$.

The terms which describe deviations from the Minkowski metric are interpreted here as fluctuations on a background. These fluctuation are assumed to be of general nature, that is, they are not restricted to be given by gravitational waves.
We consider two moments (the mean and the variance) to fully characterize the underlying fluctuation process. 

Furthermore, inspired by the analysis in \cite{DSR1PRD2006} we assume that every particle - quantum mechanically characterized by a wave packet -  has its own finite space--time resolution scale $(\Delta t_p, V_p)$. 
If the space--time fluctuations are of very long wavelength and low frequency, then the $h_{\mu\nu}$ are nearly constant and, thus, can be absorbed as constant phase into the wave function leading to no effect. For space--time fluctuations of very short wavelengths and high frequency the interaction with a quantum particle which is represented by the wave packet is given by the averaging procedure over the space--time interval $(\Delta t_p, V_p)$.
\begin{equation}
\langle h_{\mu\nu}(\mathbf{x},t)\rangle = \gamma_{\mu\nu}, \quad \delta_{\rho\sigma} \langle h^{\rho\mu}(\mathbf{x}, t) h^{\sigma\nu}(\mathbf{x}, t) \rangle  = \sigma^2_{\mu\nu}, \label{average1}
\end{equation}
where we understand the brackets as the average over background space--time. 
From now on results obtained by calculation of the space--time average $\langle \cdots \rangle$ are restricted to short wavelength and high frequency space--time fluctuations which scale is shorter than the resolution scale $(\Delta t_p, V_p)$. Possible intermediate wavelengths and frequencies of fluctuations have to be treated separably. As a result of this approach each particle can only detect averaged space--time fluctuations over the scale $(\Delta t_p, V_p)$. This can be understood as a coarse-graining of space--time fluctuations. 

Here the mean $\gamma_{\mu\nu}$ may also be interpreted as a background field, that is, as $g^{(0)}_{\mu\nu}$ in a decomposition 
\begin{equation}
g_{\mu\nu} = g^{(0)}_{\mu\nu} + h_{\mu\nu} \label{StrongBackground}
\end{equation}
(compare, e.g., \cite{MTW73}). For simplicity, we assume only a Newtonian background field and take this Newtonian gravitational field as part of the fluctuations which does not average to zero. In the case of a strongly curved background one should continue with (\ref{StrongBackground}) instead of (\ref{modmetric1}). 

\section{Modified quantum dynamics}

We start with the action for a scalar field $\phi \equiv \phi(\mathbf{x},t)$ \cite{BirDav1982} choosing the minimally coupled case
\begin{equation}
 S=\frac{1}{2}\int d^4x \sqrt{g} \left(g^{\mu\nu} \partial_\mu\phi^* \partial_\nu\phi - \frac{m^2c^2}{\hbar^2} \phi^* \phi\right) \, ,
\end{equation}
where $g=-{\rm det }[g_{\mu\nu}]$ is the determinant of the metric. Variation of the action yields the minimally coupled Klein-Gordon equation
\begin{eqnarray}
g^{\mu\nu}D_{\mu}\partial_{\nu}\phi-\frac{m^2c^2}{\hbar^2}\phi=0,
\end{eqnarray}
where $D_{\mu}$ is the covariant derivative.
We write down the covariant derivative explicitly which allows us to express the Klein-Gordon equation according to 
\begin{eqnarray}
 \partial_{\mu}g^{\mu\nu}\partial_{\nu}\phi +g^{\mu\nu}\left(\partial_{\mu}\partial_{\nu}+\frac{1}{2}\partial_\mu\ln{g} \partial_{\nu}\right)\phi-\frac{m^2c^2}{\hbar^2}\phi=0. 
\end{eqnarray}
Performing a 3+1 decomposition leads us to (latin indices $i,j$ run from 1 to 3)
\begin{eqnarray}
 0&=& g^{00}\left(\partial^2_0+\frac{1}{2}\partial_0\ln{g}\partial_0\right)\phi +\partial_0 g^{00}\partial_0 \phi + \left(\partial_0 g^{0j}\partial_j+\partial_i g^{i0}\partial_0\right)\phi \nonumber \\
& &+ g^{0j}\left(2 \partial_0\partial_j+\frac{1}{2}\partial_0\ln{g}\partial_j+\frac{1}{2}\partial_j\ln{g}\partial_0 \right)\phi +\partial_i g^{ij}\partial_j \phi\nonumber \\
& &+g^{ij}\left(\partial_i\partial_j+ \frac{1}{2}\partial_i\ln{g} \partial_j\right)\phi -\frac{m^2c^2}{\hbar^2}\phi,
\end{eqnarray}
Insertion of our metric expansion (\ref{modmetric1}) and approximating the logarithmic terms $\ln{g}\approx h-\frac{1}{2}\tilde{h}$ leads to a Klein--Gordon equation in curved space--time up to second order in the perturbations, 
\begin{eqnarray}
 0= & &\eta^{00} \left(\partial^2_0+\frac{1}{2}\partial_0\left (h-\frac{1}{2}\tilde{h}\right)\partial_0\right)\phi-h^{00}\left(\partial^2_0+\frac{1}{2}\partial_0 h\partial_0\right)\phi+\tilde{h}^{00}\partial^2_0\phi\nonumber \\
& & -h^{0i}\left(2\partial_0\partial_i+\frac{1}{2}\partial_0 h\partial_i+\frac{1}{2}\partial_i h\partial_0 \right)\phi +2\tilde{h}^{0i}\partial_0 \partial_i \phi\nonumber \\
& & + \left(\partial_0 \left(\tilde{h}^{0i}-h^{0i}\right)\partial_i+\partial_j \left(\tilde{h}^{0i}-h^{0i}\right)\partial_0\right)\phi+\partial_i\left(\tilde{h}^{ij}-h^{ij}\right)\partial_j \phi \nonumber \\
& &+\delta^{ij}\left(\partial_i\partial_j+ \frac{1}{2}\partial_i\left(h-\frac{1}{2}\tilde{h}\right) \partial_j\right)\phi
-h^{ij}\left(\partial_i\partial_j+ \frac{1}{2}\partial_i h \partial_j\right)\phi \nonumber \\
& &+\tilde{h}^{ij}\partial_i \partial_j \phi -\frac{m^2c^2}{\hbar^2}\phi,
\end{eqnarray}
where $h=\eta^{\mu\nu}h_{\mu\nu}$, $\tilde{h}=\eta^{\mu\nu}\tilde{h}_{\mu\nu}$. 

For a better interpretation, we calculate the nonrelativistic limit of this equation according to the scheme worked out by Kiefer and Singh \cite{KieferSinghPRD1991}. For doing so, we expand the phase in the wave function $\phi(\mathbf{x},t) = e^{iS(\mathbf{x},t)/\hbar}$ in powers of $c^2$
\begin{equation}
S(\mathbf{x},t)=S_0(\mathbf{x},t) c^2 + S_1(\mathbf{x},t) + S_2(\mathbf{x},t) c^{-2} + \ldots \nonumber
\end{equation}
We also expand the metric (\ref{modmetric1}) in powers of $c$
\begin{eqnarray}
h^{00} & = & c^{-2}h_{(0)}^{00}-c^{-4}h_{(1)}^{00} \nonumber\\
h^{0i} & = & c^{-1}h_{(1)}^{0i} + c^{-3}h_{(2)}^{0i}\nonumber\\
h^{ij} & = & h_{(0)}^{ij}+c^{-2}h_{(1)}^{ij} \, . \label{pertmetric}
\end{eqnarray}
We collect equal powers of the expansion parameter $c^2$ and set the resulting coefficients to zero. Starting with order $c^{4}$ we obtain 
\begin{eqnarray}
0=\left(\eta^{ij}-h_{(0)}^{ij}+\tilde{h}_{(0)}^{ij}\right)\partial_iS_0\partial_jS_0 \, .
\end{eqnarray}
As a consequence, $S_0$ can be a function of time only, $S_0=S_0(t)$.

To order $c^2$ one finds (choosing positive energy)
\begin{eqnarray}
\partial_tS_0&=& -m.\label{dtS01}
\end{eqnarray}
The next order $c^0$ yields  the nonrelativistic Schr\"odinger equation for $\psi(\mathbf{x},t) \equiv e^{iS_1(\mathbf{x},t)/\hbar}$
\begin{eqnarray}
i\hbar\partial_t\psi & = & - \frac{\hbar^2}{2m}\left(\left(\delta^{ij}-h_{(0)}^{ij}+\tilde{h}_{(0)}^{ij}\right)\partial_i\partial_j\psi+\partial_i \left(\tilde{h}_{(0)}^{ij}-h_{(0)}^{ij}\right)\partial_j \psi\right) \label{modschroed0}\\
& & - \frac{\hbar^2}{4m}\left(\delta^{ij}-h_{(0)}^{ij}\right)\partial_i{\rm Tr } h_{(0)}\partial_j\psi
+ \frac{\hbar^2}{8m}\delta^{ij}\partial_i{\rm Tr } \tilde{h}_{(0)}\partial_j\psi\nonumber\\
& & + \frac{m}{2}\left(\tilde{h}_{(0)}^{00}-h_{(0)}^{00}\right)\psi - \frac{i\hbar}{4}\partial_t \left({\rm Tr } h_{(0)}-\frac{1}{2}{\rm Tr } \tilde{h}_{(0)}\right)\psi \nonumber \\
& &+\frac{1}{2}\left\lbrace i\hbar\partial_i, h_{(1)}^{i0}-\tilde{h}_{(1)}^{i0}\right\rbrace \psi+\frac{i\hbar}{4}h_{(1)}^{i0}\partial_i {\rm Tr } h_{(0)}\psi,
\end{eqnarray}
where we introduced the following abbreviations $\tilde{h}_{(0)}^{ij}=\delta_{lm}h_{(0)}^{il}h_{(0)}^{mj}$, $\tilde{h}_{(0)}^{00}=\delta_{ij} h_{(1)}^{0i}h_{(1)}^{0j}$, $\tilde{h}_{(1)}^{i0}=-\delta_{lm} h_{(0)}^{il}h_{(1)}^{m0} $, ${\rm Tr } h_{(0)}=\delta_{ij}h_{(0)}^{ij}$ and ${\rm Tr } \tilde{h}_{(0)}=\delta_{ij}\delta_{lm}h_{(0)}^{il}h_{(0)}^{jm}$ and $\lbrace,\rbrace$ is the anticommutator.

In order to identify the hermitian parts we rewrite the Hamiltonian in a manifest covariant form by replacing the derivatives with the Laplace-Beltrami operator in the three-hypersurface,
\begin{equation}
 \Delta_{{\rm cov}}:= \frac{1}{\sqrt{^{(3)}g}}\partial_i\left(\sqrt{^{(3)}g}g^{ij}\partial_j \psi\right),
\end{equation}
where $^{(3)}g$ is the determinant of the metric $g_{ij}$. This yields to second order
\begin{eqnarray}
\Delta_{{\rm cov}}& =& \left(\delta^{ij}-h_{(0)}^{ij}+\tilde{h}_{(0)}^{ij}\right)\partial_i\partial_j +\partial_i \left(\tilde{h}_{(0)}^{ij}-h_{(0)}^{ij}\right)\partial_j \nonumber \\
& & +\frac{1}{2}\left(\delta^{ij}-h_{(0)}^{ij}\right)\partial_i {\rm Tr } h_{(0)}\partial_j - \frac{1}{4}\delta^{ij}\partial_i {\rm Tr } \tilde h_{(0)}\partial_j \, ,
\end{eqnarray}
Insertion of this operator leads to
\begin{eqnarray}
i\hbar\partial_t\psi & = &  - \frac{\hbar^2}{2m} \Delta_{{\rm cov}} \psi + \frac{m}{2}\left(\tilde{h}_{(0)}^{00}-h_{(0)}^{00}\right)\psi
+\frac{1}{2}\left\lbrace i\hbar \partial_i, h_{(1)}^{i0}-\tilde{h}_{(1)}^{i0}\right\rbrace \psi \nonumber \\
& &- \frac{i\hbar}{4}\partial_t \left({\rm Tr } h_{(0)} - \frac{1}{2} {\rm Tr } \tilde{h}_{(0)}\right)\psi + \frac{i\hbar}{4} h_{(1)}^{i0}\partial_i {\rm Tr } h_{(0)}\psi \, . \label{modhamil1}
\end{eqnarray}
Apparently, the Hamiltonian contains two time dependent nonhermitian terms. But these are nonhermitian with respect to the standard scalar product $ \left\langle \psi_1, \psi_2 \right\rangle = \int d^3x \psi_1^* \psi_2$. One has to take into account that the introduction of the metrical perturbation terms also modifies the scalar product.
Alternatively, one can transform the Hamilton operator $\mathbf{H}$ simultaneously with the wavefunction $\psi$ in order to express hermitian properties by means of the standard "flat" scalar product \cite{LamPhysLetA1995}  
\begin{eqnarray}
\psi\rightarrow \psi' = A\psi, \quad \mathbf{H} \rightarrow \mathbf{H'}=A\mathbf{H} A^{-1}+i\hbar \partial_t \ln{A}.
\end{eqnarray}
Then the scalar product reads 
\begin{eqnarray}
 \left<\psi'_1, \psi'_2 \right> = \int_V d^3x \psi'^{*}_1\psi'_2.
\end{eqnarray}
Thus the Hamiltonian can be brought to (flat) hermitian form without change of physical statements.
In our case the transformation of the wavefunction and the Hamilton operator is given by 
\begin{eqnarray}
\psi' = (^{(3)}g)^{1/4}\psi, \quad \mathbf{H'}= (^{(3)}g)^{1/4}\mathbf{H}(^{(3)}g)^{-1/4}+\frac{i\hbar}{4} \partial_t \ln{(^{(3)}g)},
\end{eqnarray}
leading to 
\begin{eqnarray}
 \mathbf{H'}\psi'&=& -(^{(3)}g)^{1/4}\frac{\hbar^2}{2m} \Delta_{{\rm cov}}\left((^{(3)}g)^{-1/4}\psi'\right) + \frac{m}{2}\left(\tilde{h}_{(0)}^{00}-h_{(0)}^{00}\right)\psi'\nonumber \\
& &+\frac{1}{2}\left\lbrace i \hbar \partial_i, h_{(1)}^{i0}-\tilde{h}_{(1)}^{i0}\right\rbrace \psi'.\label{hamtrans}
\end{eqnarray}
This Hamiltonian now is manifest hermitian with respect to the 'flat' scalar product. 

\section{The effective Schr\"odinger equation}

According to (\ref{average1}) the particle described by the Schr\"odinger equation averages over the space--time fluctuations.
For doing so, we need the averages of the fluctuating perturbation terms of the metric.
We assume that different components of the fluctuating metric are independent of each other. Then terms $\tilde{h}_{(1)}^{i0}$ which include cross terms (different pairs of indices) vanish on average and only the quadratic terms $\tilde{h}^{\mu\nu}$ effectively contribute to the modified Schr\"odinger equation. We define and note for later use 
\begin{eqnarray}
\langle h_{(0)}^{00} \rangle&=&2U(\mathbf{x}),\\
\langle \tilde{h}_{(0)}^{00}\rangle&=& \langle\delta_{lm}h_{(1)}^{0m}h_{(1)}^{0l} \rangle=\sigma_{00}^2,\\
\langle \tilde{h}_{(1)}^{i0}\rangle&=& \langle -\delta_{lm} h_{(1)}^{0m}h_{(0)}^{li} \rangle=0,\\
\langle\tilde{h}_{(0)}^{ij}\rangle&=& \langle \delta_{lm} h_{(0)}^{lj}h_{(0)}^{im} \rangle =\delta_{ij}\sigma^2_{ii} \, , \label{variance}
\end{eqnarray}
where we only kept contributions to lowest order in $c$ and where $U=U(\mathbf{x})$ is the Newtonian potential. Note that co- and contravariant averaged quantities are identical and that no summation is carried out over repeated indices $i,j$. Furthermore, we have not specified any averaging scheme.

\subsection{Spatial average}

In the following we take the spatial average of the Schr\"odinger equation (\ref{hamtrans}) by means of
\begin{equation}
\left<\ldots\right>_{V_p} := \frac{1}{V_p}\int_{V_p} d^3x \ldots \, ,
\end{equation}
where the volume $V_p=\lambda_p ^3$ depends on the particle under consideration .
That means, we have to calculate
\begin{equation}
\left\langle i \hbar \partial_t \psi \right\rangle_{V_p} = \left\langle \mathbf{H} \psi \right\rangle_{V_p} \, . \label{effDynamics1}
\end{equation}
Here we encounter expressions of the type
\begin{eqnarray}
 \left< A[h](\mathbf{x},t)D\psi(\mathbf{x},t)\right>_{V_p}&=&\frac{1}{V_p}\int_{V_p} d^3x A[h](\mathbf{x},t)D\psi(\mathbf{x},t),
\end{eqnarray}
where $A[h]$ is some function depending on the metric perturbation $h_{\mu\nu}$ and $D$ is a differential operator.
The wavefunction $\psi$ varies weakly in space over the averaging volume $V_p$ (compared to the wavelength of the fluctuations) and can therefore be decoupled from the spatial integration of the fluctuations,  
\begin{eqnarray}
\left< A[h](\mathbf{x},t)D\psi(\mathbf{x},t)\right>_{V_p}= \frac{1}{V_p}\left(\int_{V_p} d^3x A[h](\mathbf{x},t)\right)D\psi(\mathbf{x},t).
\end{eqnarray}
The averaged kinetic term $\left<(^{(3)}g^{1/4})\Delta_{{\rm cov}}(^{(3)}g^{-1/4})\psi\right>_{V_p}$ reads
\begin{eqnarray}
\left<g^{ij}\left((^{(3)}g^{1/4})\partial_i\partial_j (^{(3)}g^{-1/4})-\frac{1}{2}\partial_i \ln{\sqrt{(^{(3)}g)}}\partial_j \ln{\sqrt{(^{(3)}g)}}\right)\right>_{V_p}\psi\nonumber \\
 -\left<\frac{1}{2}\partial_i g^{ij}\partial_j \ln{\sqrt{(^{(3)}g)}}- g^{ij}\partial_i\partial_j-\partial_ig^{ij}\partial_j\right>_{V_p}\psi.
\end{eqnarray}
The first part of the covariant Laplace operator, 
\begin{eqnarray}
\left<g^{ij}\left((^{(3)}g^{1/4})\partial_i\partial_j (^{(3)}g^{-1/4})-\frac{1}{2}\partial_i \ln{\sqrt{(^{(3)}g)}}\partial_j \ln{\sqrt{(^{(3)}g)}}\right)\right>_{V_p}\psi\nonumber \\
-\left<\frac{1}{2}\partial_i g^{ij}\partial_j \ln{\sqrt{(^{(3)}g)}}\right>_{V_p}\psi, 
\end{eqnarray}
is constant (in space) dependent on quadratic terms of the fluctuations, since linear terms vanish on average. This yields an additive term $\alpha_0(t)$ in the Hamiltonian.
The second part yields
\begin{eqnarray}
\left< g^{ij}\partial_i \partial_j + \partial_i g^{ij}\partial_j\right>_{V_p}\psi & = &\left< (\delta^{ij}-h_{(0)}^{ij}+\tilde{h}_{(0)}^{ij})\partial_i  + \partial_i(\tilde{h}_{(0)}^{ij}-h_{(0)}^{ij})\right>_{V_p}\partial_j \psi \nonumber \\
&=& (\delta^{ij}+\alpha^{ij}(t) ) \partial_i \partial_j\psi + \delta ^{ij}\beta_i(h^2)\partial_j \psi,
\end{eqnarray}
where $\alpha^{ij}(t) = \langle\tilde{h}_{(0)}^{ij}(\mathbf{x},t)-h_{(0)}^{ij}(\mathbf{x},t)\rangle_{V_p}$.
\\ Finally the spatial average of the last term in (\ref{hamtrans}) reads
\begin{eqnarray}
i\hbar\left<(h_{(1)}^{i0}-\tilde{h}_{(1)}^{i0})\partial_i \psi+\frac{1}{2}\partial_i(h_{(1)}^{i0}-\tilde{h}_{(1)}^{i0})\psi \right>_{V_p}.
\end{eqnarray}
Linear perturbations are neglected again and both terms can be factorized. By virtue of the definitions of the averages of the fluctuating quantities, the average of $\tilde{h}_{(1)}^{i0}$ vanishes (no cross--correlations) and consequently the first term. The second term yields a spatially constant but time--dependent perturbation term which we will absorb into $\alpha_0(t)$.
This is also true for the last (potential-) term in the Hamiltonian yielding
\begin{equation}
 \frac{m}{2}\left<\tilde{h}_{(0)}^{00}-h_{(0)}^{00}\right>_{V_p}\psi=-mU(\mathbf{x})\psi +\frac{m}{2}\sigma^2_{00}(t)\psi.
\end{equation}
Effectively we get the spatially averaged Schr\"odinger equation 
\begin{eqnarray}
i\hbar \partial_t \psi(\mathbf{x},t) & = & - \frac{\hbar^2}{2m}\left(\delta^{ij}+\alpha^{ij}(t)\right) \partial_i\partial_j\psi(\mathbf{x},t)- \frac{\hbar^2}{2m}\delta^{ij}\beta_i(h^2) \partial_j \psi(\mathbf{x},t)\nonumber \\
& & +\alpha_0(t)\psi(\mathbf{x},t)-m U(\mathbf{x})\psi(\mathbf{x},t),
\end{eqnarray}
where the averaged quantities are time dependent.
It is possible to find a transformation under which the term $\delta^{ij}\beta_i(h^2) \partial_j$ vanishes, therefore we can omit this term and obtain finally
\begin{eqnarray}
i\hbar \partial_t \psi(\mathbf{x},t) = \left(\mathbf{H}_0+\mathbf{H}_p\right)\psi(\mathbf{x},t) \, ,\label{modschroed2}
\end{eqnarray}
where we introduced the unperturbed Hamiltonian 
\begin{equation}
\mathbf{H}_0 = \frac{-\hbar^2}{2m} \nabla^2 - m U \label{unperturbedHamiltonian}
\end{equation}
and the perturbation term 
\begin{equation}
\mathbf{H}_p (t')=\frac{-\hbar^2}{2m}\left(\alpha^{ij}(t)\partial_i \partial_j+\alpha_0(t)\right)\, , \label{perturbationterm}
\end{equation}
which incorporates the time--dependent fluctuation part.

We introduced the metric fluctuations without specifying explicitly its diagonal elements -- in general they are distinct from each other. This allows to describe an anisotropy of space and we will assume that this is the outcome of space--time fluctuations. Therefore we regard the terms $\alpha^{ij}(t)$ as an anomalous  -- still time--dependent -- inertial mass tensor which was extensively studied by Haugan \cite{Haugan1979AnnPhys}. Consequently, this quantity violates rotational invariance. Thus we can conclude that in our approach the information about space--time fluctuations is encoded in the anomalous inertial mass tensor, generally breaking rotational invariance \footnote{This is not necessarily restricted to our model of space--time fluctuations as one can introduce anisotropic gravitational waves which also lead to the violation of rotational invariance. For example stochastic gravitational waves generated by unresolved binary systems constitute an anisotropic galactic background \cite{SchutzCQG1999}.}. 

\subsection{Time--evolution}

The spatially averaged fluctuation terms and, thus, the rescaled inertial mass still fluctuate in time. Since we assumed high--frequency fluctuations it is reasonable to argue that in an experiment the fluctuations are effectively smeared out depending on the particle--properties. These are characterized by a particle dependent averaging time $\Delta t_p$ which appears in the exponent of the unitary time--evolution operator. Quadratic quantities of the metrical fluctuations are the surviving contributions.
The time evolution of the wavefunction is generated by the time--evolution operator $\mathbf{U}$, which in our case is given by
\begin{equation}
\mathbf{U}(t,0) = \exp{\left(- \frac{i}{\hbar} \mathbf{H}_0 t\right)}\exp{\left(- \frac{i}{\hbar} \int^{t}_{0} dt' \mathbf{H}_p (t')\right)} \, , \label{U0}
\end{equation}
where time--ordering is understood and the commutator yields $[\mathbf{H}_0,\mathbf{H}_p (t')]=0$. 
Now we split the perturbation term into a positive definite, constant part $\tilde{\alpha}$ and a non-definite operator $ \gamma(t)$ via $\int^t_0 dt'\mathbf{H}_p (t')=\tilde{\alpha}t+ \gamma(t)$, where the fluctuating part $\gamma(t)$ vanishes in the temporal average. This expression could in principle be generalized in terms of a polynomial expansion of the type $\sum^n_{i=1}=\alpha_i t^i$. The higher order terms ($i>1$) would render a time--dependent kinetic energy which could take any value (for long times). In order to prevent the particle gaining infinite energy out of spacetime fluctuations (or to loose all its energy to these fluctuations and, thus, effectively disappear at some stage), the kinetic energy is assumed to give a fluctuation induced effective linear growth of the phase and, equivalently, a constant $\tilde{\alpha}$. 
Because we assumed high--frequency fluctuations, which act on a timescale much shorter than the considered evolution time $t > \Delta t_p$, this allows us to understand the integral of the positive--definite part $\tilde{\alpha}$ as an average over time $t$. Now equation (\ref{U0}) reads
\begin{equation}
\mathbf{U}(t,0) = \exp{\left(- \frac{i}{\hbar} \left(\mathbf{H}_0+ \tilde{\alpha}\right)t \right)} \exp{\left(- \frac{i}{\hbar} \gamma (t)\right)}.
\end{equation} 

Finally, this yields an effective Schr\"odinger equation by taking the time derivative of the time--evolution operator and we get
\begin{equation}
 i\hbar\partial_t \psi(\mathbf{x},t)=-\frac{\hbar^2}{2m}\left(\delta^{ij}+\tilde{\alpha}^{ij}\right)\partial_i \partial_j \psi(\mathbf{x},t) -m U(\mathbf{x})\psi(\mathbf{x},t) \, , \label{effhamiltonian}
\end{equation}
where the fluctuating phase $\gamma(t)$ is absorbed into the wave function which is still denoted by the same symbol. In interferometry experiments this fluctuating phase may lead to a decrease of the visibility of the interference pattern. Theoretically speaking, this is the main result of our paper.

\section{Implications on a modified inertial mass}

In analogy to the analysis of fluctuations as function of time carried through in \cite{Radeka1988} we can associate a spectral density $S(\mathbf{k})$ with the variance in (\ref{variance}), 
\begin{equation}
(\sigma^2)^{ij} = \frac{1}{V_p} \int_{V_p} d^3x \tilde{h}_{(0)}^{ij}(\mathbf{x},t) = \frac{1}{V_p} \int_{(1/\lambda_p)^3} d^3k  (S^2(\mathbf{k},t))^{ij} \, ,
\end{equation}
which allows us - in principle - to calculate the metric fluctuation components $\tilde{\alpha}^{ij}$. 

A particular class of models is given by a power law spectral noise density $(S^2(\mathbf{k},t))^{ij} = (S_{0n}^2)^{ij} |\mathbf{k}|^n$ where $S_{0n}$ is some constant of dimension $({\rm length})^{3+n/2}$. In this case we obtain $(\sigma^2)^{ij} = (S_{0n}^2)^{ij} \lambda_p^{- (6 + n)}$. Since the fluctuations under consideration should originate from a quantum gravity scenario, we assume that $S_{0n} \approx l_{\rm Pl}^{3+n/2}$, where $l_{\rm Pl}$ is the Planck length. As a result we then obtain $(\sigma^2)^{ij} \approx \left(l_{\rm Pl}/\lambda_p\right)^{\beta}a^{ij}$, where $\beta=6+n$ and $a^{ij}$ is a tensor of order $\mathcal{O}(1)$ being diagonal and reflecting the anisotropic behaviour.
A special model can now be applied in which the outcome of any experiment measuring length intervals is influenced by spacetime fluctuations such that perturbations of this quantity scales like $\left(l_{\rm Pl}/l\right)^{\beta}$ \cite{GAC2000PRD,JackNg2003,JackNg2002}, where $\beta\geq 0$. As pointed out in \cite{JackNg2003}, common values for $\beta$ are $1,2/3,1/2$.
Since $\tilde{\alpha}^{ij}$ is positive--definite and the fluctuation of the metric as calculated in \cite{JackNg2003}, too, we apply this model of spacetime fluctuations to our scheme via the identification $\tilde{\alpha}^{ij}=\left(l_{\rm Pl}/\lambda_p \right)^{\beta}a^{ij}$. As a consequence, the modified kinetic term of the averaged Hamilton operator of equation (\ref{modschroed2}) reads
\begin{equation}
\mathbf{H}_{{\rm kin.}}=-\frac{\hbar^2}{2m}\left(\delta^{ij}+\left(\frac{l_{\rm Pl}}{\lambda_p}\right)^{\beta}a^{ij}\right)\partial_i \partial_j  \, . \label{flucscenario}
\end{equation}

In our analysis of the space--time average we encountered an anomalous inertial mass tensor which can be understood as a modification of the inertial mass - namely it is subject to a renormalization. By virtue of this modification the inertial mass affected by space--time fluctuations can be rewritten as a rescaled inertial mass which is the physical quantity measurable in experiments.
The modified inertial mass in equation (\ref{effhamiltonian}) is identified by 
\begin{equation}
(\tilde{m}^{ij})^{-1}\equiv \frac{1}{m}\left(\delta ^{ij}+ \tilde{\alpha}^{ij}\right),
\end{equation}
what can alternatively be rewritten as 
\begin{equation}
\tilde{m}_i\equiv m \cdot (1+ \alpha_p^i)^{-1},
\end{equation}
where $\tilde{\alpha}^{ij}={\rm diag }\left[\alpha_p^1, \alpha_ p^2, \alpha_p^3\right]$ and $\alpha_p^i=\delta^{ij}\left<\tilde{h}_{(0)}^{ii}\right>_{V_p}$. The index $p$ stands for ''particle'' characterized by its spatial resolution scale $\lambda_p$.
Note that the dependency of the anisotropic inertial mass on only one index ($i$) is a result of the fact that $\tilde{\alpha}^{ij}$ is diagonal.
All space--time dependent fluctuation properties of the inertial mass are now encoded in the quantities $\alpha^i$ which are proportional to the averaged quadratic perturbations $\sigma^2_{ii}$.
This gives for our fluctuation scenario (\ref{flucscenario})
\begin{equation}
\tilde{m}_{i}=\frac{m}{1+(l_{\rm Pl}/\lambda_p)^{\beta} a^{ii}}.
\end{equation}

The modified inertial mass $\tilde{m}_i$ is now a function of the fluctuation scenario $\beta$ and of the resolution scale $\lambda_p$, hence the modification is particle dependent. This modification becomes larger for increasing mass $m$. The validity of the observations made here is restricted to coherent systems, since we are dealing with modifications of quantum mechanics. Therefore one may think of large quantum systems, e.g. Bose Einstein condensates (BECs) as a favoured probe for experimental tests of our fluctuation scenario which, however, needs further investigations. Being a characteristic property of every particle species we use the Compton--length $\lambda_C = \hbar/(mc)$ as averaging scale which leads to definite estimates on fluctuation effects. This will allow us to give some estimates for possible violations of standard physics.

\section{Violation of weak equivalence principle}

By means of the rescaled inertial mass and the dependency of the modification on the particle type $\lambda_p$ one expects that the weak equivalence principle (WEP) is violated. We identify $ \tilde{m}_j=m_{\rm i}$ and $ m=m_{\rm g}$ as inertial and gravitational mass, thus the ratio of both masses yields
\begin{equation}
\left(\frac{m_{\rm g}}{m_{\rm i}}\right)^{i}_p = 1 + \alpha^{i}_p \, .
\end{equation}
This expression renders an apparent violation of the weak equivalence principle. 
In terms of the fluctuation scenario (\ref{flucscenario}) the violation factor $\alpha_p$ yields
\begin{equation}
\alpha^i_p=\left(\frac{l_{\rm Pl}}{\lambda_p}\right)^{\beta}a^{ii} \, .
\end{equation}
In order to obtain some estimates, we set the resolution scale to the Compton length $\lambda_{pj}=\hbar/(m_j c)$. For Cesium this gives $\lambda_{\rm Cs}\approx 10^{-18}\;{\rm m}$ and for Hydrogen we get $\lambda_{\rm H}\approx 10^{-16}\;{\rm m}$. Assuming rotational invariance ($\alpha^i_p=\alpha_p $) we get
\begin{equation}
\alpha_{\rm Cs}\approx\left(\frac{10^{-35}} {10^{-18}}\right)^{\beta}=(10^{-17})^{\beta}\, ,  \quad \alpha_{\rm H}\approx\left(\frac{10^{-35}} {10^{-16}}\right)^{\beta}=(10^{-19})^{\beta}\, .
\end{equation}
For the calculation of the E\"otv\"os factor we need the accelerations, namely
\begin{equation}
a_{\rm Cs}=(1+\alpha_{\rm Cs})g=(1+(10^{-17})^{\beta})g, \quad a_{\rm H} = (1 + \alpha_{\rm H})g=(1+(10^{-19})^{\beta}) g.
\end{equation}
The E\"otv\"os factor reads  
\begin{eqnarray}
\eta = 2\frac{|a_{\rm Cs} - a_{\rm H}|}{|a_{\rm Cs} + a_{\rm H}|} 
\approx |\alpha_{\rm Cs} - \alpha_{\rm H}| \approx (10^{-17})^{\beta}.
\end{eqnarray}
In our model this yields
\begin{eqnarray}
\eta_{\beta=1}=10^{-17}, \quad \eta_{\beta=2/3}=10^{-12}, \quad \eta_{\beta=1/2}=10^{-9}.
\end{eqnarray}
Regarding the precision of current atom interferometers \cite{PetChungChu1999} this would rule out the random--walk scenario $\beta=1/2$, but one has to take into account that our scenario is quite optimistic, regarding the amplitude of fluctuations estimated by the spatial resolution scale $\lambda_p=\lambda_C$. Nevertheless, it suggests that violations of the weak equivalence principle should be tested with high--precision atomic interferometry in space missions like GAUGE \cite{GAUGEMEETING2007} and HYPER \cite{HYPERESA}, which may provide improved sensitivities capable of ruling out certain fluctuation scenarios. Furthermore, the excellence cluster QUEST \cite{QUEST} provides opportunities for further development of high--precision experiments and interdisciplinary collaborations, which will press ahead the search for this kind of violation scenarios. 

\section{Conclusions}

In this work we discussed the influence of space--time fluctuations - interpreted as classical fluctuations of the metric. We assumed that these fluctuations are anisotropic, realized by distinct diagonal elements of the perturbation terms. This breaks rotational invariance and an anomalous inertial mass tensor appears in the effective Schr\"odinger equation. 
Assuming high frequency- and small wavelength space--time fluctuations gives rise to a modified inertial mass after having calculated the space--time average.
We have seen, that the modification of the inertial mass is dependent on the type of particle which is an effect of the finite resolution scale of the particle $(\lambda_p)$ and leads to an apparent violation of the weak equivalence principle of the order $\mathcal{O}(10^{-17}-10^{-12})$, being dependent on the fluctuation scenario. This effect may be explored by Bose--Einstein condensates and by matter--wave interferometry. Generally, our results which are dependent on the resolution scale are also viable in the case where rotational invariance is still valid.
Thus our model incorporates several effects which are possible outcomes of a theory of quantum gravity.

\section*{Acknowledgments}
It is a pleasure to thank A. Camacho, H. Dittus, F. Klinkhamer, A. Macias, E. Rasel, and C. Wang for discussions. We also like to thank the members of the cluster of excellence QUEST (Quantum Engineering and Space--Time Research) for the fruitful collaboration. This work has been supported by the German Aerospace Center (DLR) grant no. 50WM0534 and the German Research Foundation (DFG).

\end{document}